\documentclass[review]{siamart1116}

\usepackage{lipsum}
\usepackage{amsfonts}
\usepackage{graphicx}
\usepackage{epstopdf}
\usepackage{algorithmic}
\ifpdf
  \DeclareGraphicsExtensions{.eps,.pdf,.png,.jpg}
\else
  \DeclareGraphicsExtensions{.eps}
\fi

\usepackage{csquotes}

\usepackage{nicefrac}

\numberwithin{theorem}{section}

\newcommand{\TheTitle}{Fractional PKS equations for chemotactic superdiffusion} 
\newcommand{\TheAuthors}{G. Estrada-Rodriguez, H. Gimperlein, and K. J. Painter}

\headers{\TheTitle}{\TheAuthors}

\title{{Fractional Patlak-Keller-Segel equations for chemotactic superdiffusion}\thanks{Submitted to the editors DATE.
\funding{H.~G.~acknowledges support by ERC Advanced Grant HARG 268105. G.~E.~R.~was supported by The Maxwell Institute Graduate School in Analysis and its
Applications, a Centre for Doctoral Training funded by the UK Engineering and Physical
Sciences Research Council (grant EP/L016508/01), the Scottish Funding Council, Heriot-Watt
University and the University of Edinburgh. K.~J.~P.~acknowledges support from the Politecnico di Torino for a Visiting Professorship award.}}}

\author{
  Gissell Estrada-Rodriguez\thanks{Maxwell Institute for Mathematical Sciences and Department of Mathematics, Heriot-–Watt University, Edinburgh, EH14 4AS, United Kingdom
    (\email{ge5@hw.ac.uk}, \email{h.gimperlein@hw.ac.uk}, \email{k.painter@hw.ac.uk}).}
  \and
  Heiko Gimperlein${}^\dag$\thanks{Institute for Mathematics, University of Paderborn, Warburger Str. 100, 33098 Paderborn, Germany.}
  \and
  Kevin J. Painter${}^\dag$\thanks{Politecnico di Torino, Dipartimento di Scienze Matematiche, Corso Duca degli Abruzzi, 24, 10129 Torino TO, Italy.}
}

\usepackage{amsopn}

\ifpdf
\hypersetup{
  pdftitle={\TheTitle},
  pdfauthor={\TheAuthors}
}
\fi

\usepackage{braket,amsfonts}

\usepackage{mathtools}

\usepackage{amsfonts}
\usepackage{dsfont}

\usepackage{array}

\usepackage[caption=false]{subfig}

\newsiamthm{claim}{Claim}
\newsiamremark{rem}{Remark}
\newsiamremark{expl}{Example}
\newsiamremark{hypothesis}{Hypothesis}
\crefname{hypothesis}{Hypothesis}{Hypotheses}

\usepackage{algorithmic}

\usepackage{graphicx,epstopdf}

\Crefname{ALC@unique}{Line}{Lines}

\usepackage{amsopn}

\numberwithin{theorem}{section}

\usepackage{xspace}
\usepackage{bold-extra}
\usepackage[most]{tcolorbox}
\definecolor{OliveGreen}{rgb}{0,0.6,0}

\colorlet{texcscolor}{blue!50!black}
\colorlet{texemcolor}{red!70!black}
\colorlet{texpreamble}{red!70!black}
\colorlet{codebackground}{black!25!white!25}

\lstdefinestyle{siamlatex}{%
  style=tcblatex,
  texcsstyle=*\color{texcscolor},
  texcsstyle=[2]\color{texemcolor},
  keywordstyle=[2]\color{texemcolor},
  moretexcs={cref,Cref,maketitle,mathcal,text,headers,email,url},
}

\tcbset{%
  colframe=black!75!white!75,
  coltitle=white,
  colback=codebackground, 
  colbacklower=white, 
  fonttitle=\bfseries,
  arc=0pt,outer arc=0pt,
  top=1pt,bottom=1pt,left=1mm,right=1mm,middle=1mm,boxsep=1mm,
  leftrule=0.3mm,rightrule=0.3mm,toprule=0.3mm,bottomrule=0.3mm,
  listing options={style=siamlatex}
}

\newtcblisting[use counter=example]{example}[2][]{%
  title={Example~\thetcbcounter: #2},#1}

\newtcbinputlisting[use counter=example]{\examplefile}[3][]{%
  title={Example~\thetcbcounter: #2},listing file={#3},#1}

\DeclareTotalTCBox{\code}{ v O{} }
{ 
  fontupper=\ttfamily\color{black},
  nobeforeafter,
  tcbox raise base,
  colback=codebackground,colframe=white,
  top=0pt,bottom=0pt,left=0mm,right=0mm,
  leftrule=0pt,rightrule=0pt,toprule=0mm,bottomrule=0mm,
  boxsep=0.5mm,
  #2}{#1}

\patchcmd\newpage{\vfil}{}{}{}
\begin{document}

\maketitle

\begin{abstract}
The long range movement of certain organisms in the presence of a chemoattractant can be governed by long distance runs, according to an approximate L{\'e}vy distribution. This article clarifies the form of biologically relevant model equations: We derive Patlak-Keller-Segel-like equations involving nonlocal, fractional Laplacians from a microscopic model for cell movement. Starting from a power-law distribution of run times, we derive a kinetic equation in which the collision term takes into account the long range behaviour of the individuals. A fractional chemotactic equation is obtained in a biologically relevant regime. Apart from chemotaxis, our work has implications for biological diffusion in numerous processes.
\end{abstract}

\begin{keywords}
  Chemotaxis, Patlak-Keller-Segel equation, velocity-jump model, nonlocal diffusion, L\'{e}vy walk, cell motility.
\end{keywords}

\begin{AMS}
   92C17, 35R11, 35Q92
\end{AMS}

\section{Introduction}

Chemotaxis is the directed movement response of a cell or organism to some chemical concentration gradient, and has been identified in areas as diverse as microbiology \cite{baker2006signal,berg1972chemotaxis,manahan2004chemoattractant}, developmental biology \cite{dormann2006chemotactic}, immunosurveillance \cite{lauffenburger1983localized,wang2011signaling}, cancer development \cite{yamaguchi2005cell} and animal movement \cite{kennedy1974pheromone,ward1973chemotaxis}.

Motivated by these applications, chemotaxis and related phenomena have received significant attention in the theoretical community, e.g. see the reviews \cite{hillen2009user,horstmann20031970}. Modelling approaches range from microscopic to macroscopic, with the early and seminal contributions of Patlak \cite{patlak1953random} and Keller and Segel \cite{keller1970initiation,keller1971model}, respectively, providing examples. Chemotactic models derived from microscopic perspectives have tended to follow 
standard assumptions on the behaviour of individuals, usually assuming that the search 
strategy follows a biased random walk. In particular, the distribution of 
times between reorientations by the cells/organisms is taken to follow a Poisson 
distribution, as backed up by observations of 
{\em E. coli} such as \cite{berg1972chemotaxis}, and the result is a Fickian-type diffusive 
flux when a continuous approximation is derived.

In this work, motivated by real world examples (discussed in \Cref{sec: examples}), we assume the motion follows a long-tailed distribution of run times. From a microscopic model for chemosensitive movement  we derive fractional Patlak-Keller-Segel equations for the density ($\bar{u}$) of some chemotactic population in the presence of a chemoattractant (of concentration $\rho$). The fractional Patlak-Keller-Segel system obtained here is 
\begin{equation}
\begin{aligned}
\partial_t\bar{u}  & =c_0\nabla\cdot(D_\alpha\nabla^{\alpha-1}\bar{u}-\chi\bar{u}\nabla\rho),\\ \partial_t\rho & =D_\rho\Delta\rho+f(\bar{u},\rho) \label{eq: model}.
\end{aligned}
\end{equation}
$\nabla^{\alpha-1}$ denotes a fractional gradient which interpolates between ballistic motion ($\alpha=1$) and ordinary diffusion ($\alpha=2$); note that the case $\alpha=2$ corresponds to the classic formulation of Patlak-Keller-Segel equations proposed phenomenologically in \cite{keller1970initiation}. The second equation is of standard reaction-diffusion form, assuming that the chemical diffusion with coefficient $D_\rho$ is not affected by the nonlocal behaviour of the organisms. The chemotactic population is governed by a diffusion term with coefficient $D_\alpha$ (defined at the end of \Cref{sec: scaling})
that represents a random component to motility, and a chemotactic flux of advective type, 
where the advection is proportional to the chemical gradient. The function 
$\chi$ is commonly referred to as the chemotactic sensitivity. In the case of constant $D_\alpha$ we obtain an honest fractional Laplacian, namely, $\nabla\cdot\nabla^{\alpha-1}=c(-\Delta)^{\nicefrac{\alpha}{2}}$ for $1<\alpha<2$. Unlike recent analyses which obtain fractional behaviour in different contexts (e.g. \cite{taylor2016fractional} and \cite{bellouquid2016kinetic}), we consider the derivation from a fully microscopic description, according to how the concentration and gradient of chemoattractant influences the movement of individual organisms.   

Starting from a velocity jump model in which an individual performs occasional long jumps according to an approximate L\'{e}vy distribution, we derive the appropriate kinetic-transport equation where the collision term describes the nonlocal motion. We then use a perturbation argument and an appropriate hyperbolic scaling in space and time, obtaining system \cref{eq: model} in the limit.

\subsection{L\'{e}vy walks and motivating examples}\label{sec: examples}

Our work is motivated by experimental results which indicate the presence of behaviour with characteristics similar to a L\'{e}vy walk as an alternative search strategy, particularly when chemoattractants, 
food or other targets are sparse or rare; examples include \cite{harris2012generalized,levandowsky1997random,li2008persistent,ramos2004levy}. 
In contrast to Brownian motion, a L\'{e}vy walk includes 
a non-negligible probability for long positional jumps. In a biological context these \enquote{long jumps} correspond 
to persisting in a single direction of motion for a substantially longer time than 
in typical random walks. 
The distribution of runs asymptotically behaves like a power-law distribution with finite mean, but unbounded variance. 
While for Brownian motion the mean squared displacement $\left<x^2\right>$ of a 
particle is a linear function of time, for a power-law distribution with power $1<\alpha<2$, $\left<x^2\right>\sim t^{\nicefrac{2}{\alpha}}$ grows faster for large times. The exponent $\alpha=1$ corresponds to ballistic transport, while $\alpha=2$ is the case of normal diffusion.  For a review of L\'{e}vy walk models and their ubiquitous appearance we refer to \cite{RevModPhys.87.483}. 

To motivate the present modelling, we describe two systems in which organisms 
with well documented chemotactic responses have been suggested to display L\'{e}vy walk 
characteristics. Moreover, we note that L\'{e}vy walk behaviour has been suggested in numerous biological contexts, e.g. immune cells \cite{harris2012generalized}, ecology \cite{bullock2017synthesis} and human populations \cite{rhee2011levy}.

\subsubsection{\textit{E. coli}}
The chemotactic behaviour of the bacterium {\em {E. coli}} has been extensively studied,
such that more is known for its signalling pathways and mechanistic control of 
chemotaxis than for any other system \cite{kollmann2005design}. Motile {\em {E. coli}} carry long 
flagella that allow them to move in ``run and tumble'' fashion: counterclockwise rotation of flagella 
results in their bundling, and smooth swimming occurs with an 
approximately fixed heading; rotation clockwise results in outward flaying, 
and the bacterium tumbles randomly while maintaining an almost fixed position. In the 
presence of a chemoattractant, rotation is controlled by a signalling pathway, where attachment of the chemical to a membrane bound receptor induces signalling to the flagellum's rotatory machinery. Chemotaxis is achieved by increasing the run 
time when the cell experiences an increasing attractant gradient, so that on
average an individual spends more time moving up gradients than down them.

While classic experiments \cite{berg1972chemotaxis} indicate that the distribution of 
tumbling events for \textit{E. coli} follows a Poisson distribution, more recent 
experiments  \cite{korobkova2004molecular} have shown that (for mediums where the attractant is absent) the distribution of runs can have a heavy tail, suggesting that the 
bacteria may follow 
a L\'{e}vy walk in particular environments. A theoretical study carried out in \cite{tu2005white} suggested that temporal fluctuations of a key protein in the signalling pathway of \textit{E. coli} can induce power-law distributions of the run times, in agreement with the previous experimental results \cite{korobkova2004molecular}. A modelling study in \cite{matthaus2009coli}
suggested that the switch from local (Brownian) to nonlocal (L\'{e}vy) search in particular depends on CheR activity (a cytoplasmic signalling protein 
regulating receptor activity). In the case of fluctuating CheR, the running behaviour of \textit{ E. coli} 
 followed a power-law distribution, while for constant CheR it followed a Brownian motion. Simulations showed that for the case of fluctuating CheR, 
bacteria subsequently found food faster as switching between long and short runs allowed them
to leave nutrient depleted patches and reinitiate searching.
\subsubsection{\textit{Dictyostelium discoideum (Dd)}}
Similar findings have been suggested in the searching strategy of certain amoeboid cells, and
in particular the cellular slime mold {\em Dd}. {\em Dd} is the classic model system for 
studying chemotaxis behaviour in eukaryotic cells, displaying a complex life cycle in which it
switches between unicellular (``vegetative swarming'', in which single cells 
migrate, consume and divide) and multicellular (cells self-organise into a 
collective of $\sim 10^5$ cells and behave as a single organism) phases. The 
chemotactic response of {\em Dd} to the chemoattractant cAMP is well known to 
control multicellular phases \cite{bonner2009social},
but chemotaxis is also a crucial mechanism for finding food during vegetative phases;
for example, chemotaxis to folic acid allows {\em Dd}
to seek out folate-secreting bacteria prey \cite{pan1975determination}.

A study of Levandowski et al. \cite{levandowsky1997random}, where a group of 17 soil amoeba of 8 different types were isolated and tracked in a medium free of nutrients, revealed that the mean 
squared displacement could follow a power-law distribution, suggesting the L{\'e}vy walk model as an approach
to describe the movement of these organisms. 
Nevertheless, the study also remarked that the duration of the experiment may not be
sufficient to see if cells switched to a normal distribution at longer times. 

\begin{figure}[htbp]
    \centering
   \includegraphics[scale=0.5]{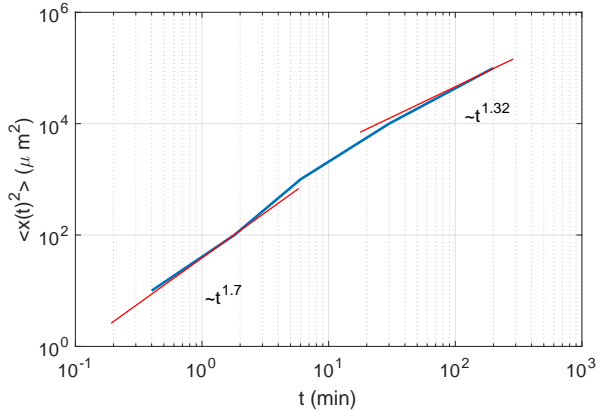}
    \caption{Reproduction of the data reported in \cite{li2008persistent}. Average of the mean squared displacement $\left<x(t)^2\right>$ of 12 cells, where $x(t)^2$ was averaged over all pairs of time points for each trajectory. As described in \cite{li2008persistent}, each cell was followed for $8-10\ \textnormal{hrs}$ with a sampling interval of $10\ \textnormal{s}$.}
    \label{fig: msd of cells}
\end{figure}

A more recent research explored the motility of \textit{Dd} and \textit{Polysphondylium palladium}  cells in a 
food free medium \cite{li2008persistent}, with authors concluding 
that cells bias their motion by remembering the last turn and subsequently performing long 
runs without changing direction for $\sim9\,\mbox{min}$. Experiments tracked 12 
cells, obtaining the trajectories for each cell at different run times (results reproduced in \Cref{fig: msd of cells}). 
For short run times ($0.4\thinspace \textnormal{min}<\tau<5\thinspace\textnormal{min}$) the trajectories appear to be almost ballistic, while for larger run times ($\tau>30\thinspace\textnormal{min}$) the trajectories lie between normal diffusion and ballistic transport, suggesting a superdiffusion-like behaviour. These results are also reflected in their measurements of the cell velocities at different run times. Characteristics of the type of motion observed in \cite{li2008persistent} suggest that these cells do not specifically perform a L\'{e}vy walk, but a form of long directionally persistent random walk.

Finally, a study on the search strategy of wild type AX3 \textit{Dictyostelium} cells in the absence of attractant \cite{van2009food} indicated that starved cells search for 
food in larger areas not by increasing their speed but by biasing towards very long runs. 
In other words, cells changed their strategy from making a very localized search to expanding the search area through persisting in their motion in a single direction.

\subsection{Outline}
This paper is organized in the following way. In \Cref{sec: Assumptions} we discuss the assumptions for the type of motion of the organism that we are modelling. In \Cref{sec: modeling equations} we derive the resulting kinetic equation, and in \Cref{sec: scaling section} we introduce the relevant scaling regime. \Cref{sec: derivation of turning operator} deals with the derivation of the \enquote{collision operator} that describes the actual dynamics of the organism or cell, and finally in \Cref{sec: scaling} we obtain the fractional Patlak-Keller-Segel equation.

\section{Model assumptions} \label{sec: Assumptions}

\begin{figure}[tbhp]
  \centering
  \subfloat[]{\label{fig:a}\includegraphics[scale=0.4]{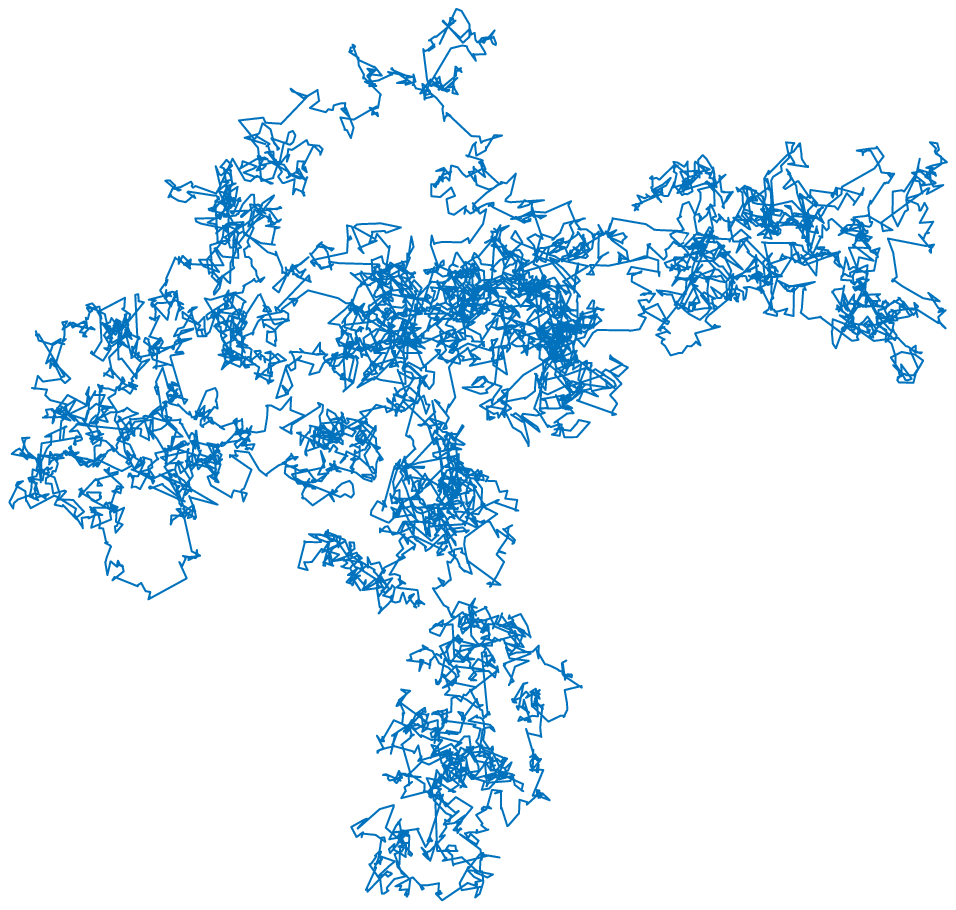}}
  \subfloat[]{\label{fig:b}\includegraphics[scale=0.4]{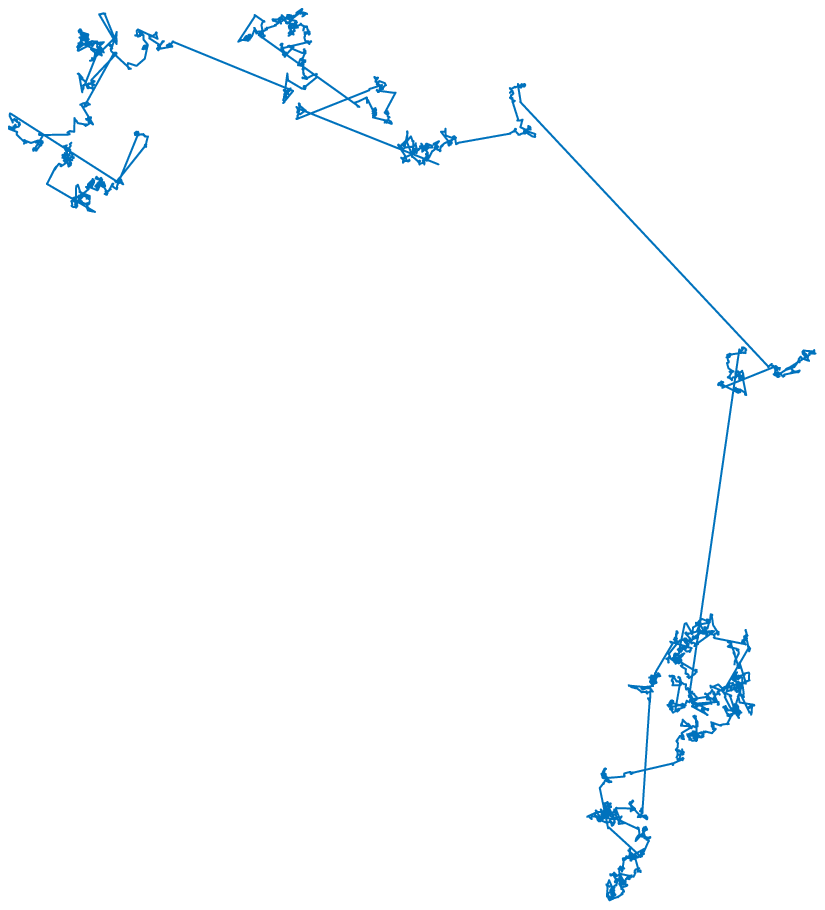}}
  \caption{Illustration of trajectories of a Brownian motion (\ref{fig:a}) and a L\'{e}vy walk (\ref{fig:b}) in two dimensions. The Brownian motion trajectory was obtained using a running probability function $\psi=e^{-\tau}$ and for the L\'{e}vy walk $\psi=\left(\frac{1}{1+\tau}\right)^\alpha$ for $\alpha=1.5$. The angle $\theta$ for the new direction is chosen from a uniform distribution in both cases.}\label{fig: Levy walk}
\end{figure}

Motivated by the experimental results in \cite{korobkova2004molecular} and \cite{li2008persistent} we model a population of organisms moving in a medium in $\mathbb{R}^n$, containing 
some chemical (with concentration $\rho=\rho(\mathbf{x},t)$) 
that acts as an attractant. We assume that each individual performs a biased random walk 
according to the distribution of $\rho$ with the following properties:
\begin{enumerate}
\item The interactions between individuals are taken to be negligible. This assumption is reasonable for the descriptions of 
experiments on {\em Dd} above (tracking spatially distributed cells) and for swimming 
{\em E. coli}, where the intracellular separation is often at least one order of 
magnitude greater than the cell diameter (e.g. \cite{berg1972chemotaxis}).
\item Starting at position $\mathbf{x}$ and time $t$, we assume an individual runs in 
direction $\theta$ for some time $\tau$, called the \enquote{run time}. Typical 
trajectories are shown in \Cref{fig: Levy walk} for different run time distributions.
\item The individuals are assumed to move with constant forward speed $c$, following a straight line motion between reorientations. 
\item Each time the individual stops it selects a new direction $\eta$ according to 
a distribution $k(\mathbf{x},t,\mathbf{\theta};\mathbf{\eta})$ which only depends on $|\theta - \eta|$. The choice 
of new direction is taken here to be independent of the chemical concentration or gradient. 
\item The reorientation is assumed to be (effectively) instantaneous.  
\item The running\footnote{In probability this is also known as survival probability, where the \enquote{event} in this case is to stop. Hence \enquote{survival} in that context refers to the probability of continuing to move in the same direction for some time $\tau$.} probability $\psi$, which is defined as the probability that an individual moving in some fixed direction does not stop until time $\tau$, is taken to depend on the environment surrounding the individual (specifically, the concentration $\rho$ and its gradient $D^\theta\rho$). Consequently, the stopping rate $\beta$ will also depend on $\rho$ and $D^\theta\rho$.
\end{enumerate}

The above assumptions are particularly relevant for the run and tumble motion of
\textit{E. coli} and similar bacteria which bias their run length according to the chemical concentration. For example, the speed $c$ of such bacteria is 
typically between $10-30\ \mu\textnormal{m/s}$ and the average length of a run is approximately ten 
times the cell body length \cite{berg2008coli, berg1972chemotaxis, macnab1972gradient}. 
Further, tumbling durations are known to be approximately $0.11\ \textnormal{s}$, an order of magnitude 
shorter than typical run times ($\sim 1.3\ \textnormal{s}$), so reorientations can be assumed to be instantaneous \cite{berg2008coli,berg1972chemotaxis}. Nevertheless, the above assumptions can be 
modified for cells like \textit{Dictyostelium} and \textit{leukocytes} \cite{alt1980biased}, under appropriate re-examination. Even the motion of larger organisms, such as butterflies, can be 
described by a persistent or correlated random walk with underlying characteristics 
similar to those described above \cite{kareiva1983analyzing,othmer1988models}.

\subsection{Turn angle distribution}

Recall that when an individual reorients in an isotropic medium, the new direction 
chosen, $\eta$, is symmetrically  distributed with respect to the previous 
direction, $\theta$ \cite{alt1980biased}. In this case 
\begin{equation}
k(\mathbf{x},t,\mathbf{\theta};\mathbf{\eta})=\ell(\mathbf{x},t,|\eta-\theta|),
\end{equation}
where $\ell$ represents a distribution and $|\eta-\theta|$ denotes the distance 
between two directions on the unit sphere $S$.

For \textit{E. coli} and certain other cells, an inhomogeneous medium
(i.e. heterogeneous chemoattractant concentration) generates a variable run length, 
such that the run length is increased when the cell experiences an increasing 
concentration of ligand. The turn angle, however, is not affected by the 
concentration \cite{berg1993random} since the bacteria are believed to be too small 
to directly sense a chemoattractant gradient \cite{endres2008accuracy}. 
Hence, during the reorientation we assume that cells choose a new direction from the distribution $\ell$, while the stopping frequency $\beta$ during the subsequent run 
is taken to depend on the ligand concentration.

We should note that in the case of larger cells, such as \textit{leukocytes} or 
\textit{Dd}, a cell can sense a chemoattractant gradient without moving (i.e. the cell is large enough to assess it directly),
and hence their next direction at a turn can also be directly influenced by the gradient. 
In this case the turn angle distribution $k$ would not be symmetric but biased according to 
the concentration and total gradient of the attractant. We do not consider this extension here.

\subsection{Running probability}

As described earlier, the motion of \textit{E. coli} depends on the 
concentration of chemoattractant via intracellular signalling 
molecules that control the tumbling phase.
As shown in \cite{korobkova2004molecular}, under certain conditions \textit{E. coli} 
can perform occasional long jumps with a corresponding power-law distribution of run 
lengths. To describe motion in such environmental conditions, we assume the following 
power distribution with exponent $\alpha$ for the running probability:
\begin{equation}
\psi(\cdot,\theta,\tau)=\left(\frac{\mathcal{S}(\rho,D^\theta \rho)}{\mathcal{S}(\rho,D^\theta \rho)+\tau}\right)^{\alpha}. \label{eq: survival}
\end{equation}
$\psi$ describes the probability that an individual running in direction $\theta$ stops after time $\tau$.
Here $\mathcal{S}(\rho,D^\theta \rho)=\tau_0(\rho)+\tau_1(\rho) D^\theta\rho$. The dot denotes dependence on space and time $(\mathbf{x},t)\in\mathbb{R}^n\times\mathbb{R}^+$.

As a remark, we note that the above choice of $\psi$ is possibly more relevant when the concentration of $\rho$ is small, i.e. when individuals need to do more searching. In 
regions of large $\rho$ it may be relevant to revert to a more classic (exponential/Poisson) choice, e.g. as in \cite{alt1980biased}: see the discussion at the end of the paper.

The running probability $\psi$ is related to the stopping frequency via 
\begin{equation}
\psi(\cdot,\theta,\tau)=\exp\left(-\int_0^{\tau} \beta(\mathbf{x}+cs\theta,t+s,\mathbf{\theta},s)ds \right).\label{eq: relation suvival and beta}
\end{equation}

This means that the probability of running for time $\tau$ without stopping is equal to the exponential of the cumulative stopping frequency.
Therefore, the stopping frequency during a run phase is given by
\begin{equation}
\beta(\cdot,\theta,\tau)=\frac{\alpha}{\tau_0+\tau_1D^\theta\rho+\tau},\label{eq:beta}
\end{equation}
in a quasi-static approximation.

\begin{rem}
As discussed in \cite{othmer2002diffusion}, the manner by which the chemoattractant 
concentration affects motion (and consequently the equations) heavily depends on the 
magnitude of the perturbation that is considered in the stopping frequency: $\beta(\cdot, \mathbf{\theta}, \tau)=\beta_0+\beta_1(\cdot, \mathbf{\theta}, \tau, D^\theta \rho)$ where $\beta_1$ is of lower order in $\varepsilon$, in the sense of \Cref{sec: scaling}. Note that here we consider that only the second term $\beta_1$ depends on $\rho$. Additive perturbations of $\psi$, e.g. 
\begin{equation}
\psi(\cdot,\theta,\tau)=(1-\varepsilon^z)\left( \frac{\varepsilon^\mu\tau_0(\rho)+\varepsilon^\mu\tau_1(\rho) D_\varepsilon^\theta\rho}{\varepsilon^\mu\tau_0(\rho)+\varepsilon^\mu\tau_1(\rho) D_\varepsilon^\theta\rho+\tau}\right)^\alpha+\varepsilon^z e^{-\tau\mathcal{C}(\rho,D^\theta_\varepsilon\rho)}, \label{eq: switching}
\end{equation}
give rise to perturbations in $\beta$ for large $\tau$ as well. For the choice \cref{eq: switching} with $z\leq 0$ we would 
also obtain a fractional Patlak-Keller-Segel type of equation as in \Cref{sec: scaling}.
\end{rem}

\section {Model equations}\label{sec: modeling equations}

We consider the assumptions from \Cref{sec: Assumptions} and are guided by Alt's
approach in \cite{alt1980biased,alt1980singular}. For a population of total density $\sigma(\mathbf{x},t,\mathbf{\theta},\tau)$, where individuals at $(\mathbf{x},t)$ move in 
direction $\mathbf{\theta}$ for some time $\tau$, the governing equations of motion are
given by
\begin{equation}
\left(\partial_{\tau}+\partial_t+c\mathbf{\theta}\cdot\nabla \right)\sigma(\cdot,\mathbf{\theta},\tau)=-\left(\beta\sigma\right)(\cdot,\mathbf{\theta},\tau),\label{eq:governing_eq1}
\end{equation}
\begin{equation}
\sigma(\cdot,\mathbf{\eta},0)=\int_{0}^{t}\int_{S}(\beta\sigma)(\cdot,\mathbf{\theta},\tau)k(\cdot,\mathbf{\theta;}\eta)d\theta d\tau.\label{eq:governing_eq2}
\end{equation}

The kinetic-transport equation \cref{eq:governing_eq1} is analogous to the Boltzmann equation, where the collision term in this case describes the behaviour of the individuals for classical velocity jump models of bacteria. It is well known that in a suitable asymptotic limit one obtains diffusion-like equations for the 
macroscopic (or  observable) density of bacteria, $$\bar{u}(\mathbf{x},t)\coloneqq\frac{1}{|S|}\int_S\int_0^t\sigma(\cdot,\theta,\tau) d\tau d\theta.$$

The left hand side of \cref{eq:governing_eq1} describes the temporal variation and transport of the 
density $\sigma$, while the right hand side gives the density of individuals
``left behind'' due to tumbling, occurring with frequency $\beta(\cdot,\mathbf{\theta},\tau)$.
The individuals that tumble undertake a reorientation process and
choose a new direction $\eta$ with probability $k(\cdot,\mathbf{\theta;}\eta)$, i.e. the 
turn angle distribution. This process is explicitly described by the initial conditions in the run time $\tau$ in 
\Cref{eq:governing_eq2}, where the left hand side is the total density of individuals 
starting a new run ($\tau=0$). This density is equal to the total population at 
$(\mathbf{x},t)$ oriented across all directions on the surface $S$ and with different 
run times $\tau$.

Note that we consider \Cref{eq:governing_eq1,eq:governing_eq2} in the whole space $\mathbb{R}^n$, thereby avoiding any specification of boundary conditions and allowing our approach to be applicable to a wide variety of systems. Further discussion of the boundary conditions is 
provided in the conclusions.\\
 
Using the method of characteristics, we can find the solution of equation \cref{eq:governing_eq1},
\begin{equation}
\sigma(\cdot,\mathbf{\theta},\tau)=\sigma(\mathbf{x}-c\mathbf{\theta}\tau,t-\tau,\mathbf{\theta},0)\exp\left(\int_{0}^{\tau}-\beta(\mathbf{x}+cs\mathbf{\theta},t+s,\mathbf{\theta},s)ds \right). \label{eq: solution}
\end{equation}

Experimentally measuring the density $\sigma$ at each $\tau$ is infeasible, and therefore we write system \cref{eq:governing_eq1}-\cref{eq:governing_eq2} in terms of a new density
\begin{equation}
\bar{\sigma}(\cdot,\mathbf{\theta})=\int_{0}^{t} \sigma(\cdot,\mathbf{\theta},\tau)d\tau, \label{eq: sigma_bar}
\end{equation}
which describes the density of the population in $\mathbf{x}$ at time $t$ and moving in direction $\theta$. Integrating over $\tau$ in \cref{eq:governing_eq1}-\cref{eq:governing_eq2} we obtain
\begin{align}
 \partial_t\bar{\sigma}+c\mathbf{\theta}\cdot \nabla \bar{\sigma} & =\sigma (\cdot,\mathbf{\theta},0)-\int_{0}^{t}(\beta \sigma)(\cdot,\mathbf{\theta},\tau)d\tau, \label{eq: 3.4}
\end{align}
where $\sigma(\mathbf{x},t,\mathbf{\theta},0)$ is analogous to \cref{eq:governing_eq2} and is given by
\begin{equation}
\sigma(\cdot,\mathbf{\theta},0)=\int_0^t\int_S(\beta\sigma)(\cdot,\eta,\tau)k(\cdot,\eta;\theta)d\eta d\tau.\label{eq: 3.6}
\end{equation}

Biologically, it is crucial that the stopping frequency $\beta$ will depend not only on the concentration 
at a given point $(\mathbf{x},t)$, but also on the gradient of the concentration along a run \cite{segel1977theoretical}, such that
$$\beta(\cdot,\mathbf{\theta},\tau)=\beta_0(\rho(\mathbf{x},t),D^\theta\rho(\mathbf{x},t)),\ \mathrm{where}\ D^\theta\rho=\partial_t \rho+c\mathbf{\theta}\cdot \nabla\rho.$$ 
This dependence of $\beta$ on $\rho$ reflects a memory process in the intracellular signalling pathway that allows the individual to assess the variation in the chemoattractant concentration along 
the run.  

The turn angle operator $T$ describes the effect of changing from direction $\mathbf{\theta}$ to a new direction $\mathbf{\eta}$. It is given by
\begin{equation}
T\phi(\eta)=\int_S k(\cdot,\mathbf{\theta};\mathbf{\eta})\phi(\theta)d\theta \label{eq: turn angle operator}.
\end{equation}

Some of its basic properties are discussed in \Cref{sec: turn_angle_properties}. Using $T$, the differential-integral equation \cref{eq: 3.4} can be re-written, with \cref{eq: 3.6}, as
\begin{align}
\partial_t\bar{\sigma}+c\mathbf{\theta}\cdot \nabla \bar{\sigma} & =\int_S k(\cdot,\mathbf{\eta};\mathbf{\theta})\int_0^{t}(\beta \sigma)(\cdot,\eta,\tau) d\tau d\mathbf{\eta}-\int_0^{t}(\beta \sigma)(\cdot,\theta,\tau) d\tau \nonumber\\
& = T\left(\int_0^{t}(\beta \sigma)(\cdot,\theta,\tau) d\tau \right) -\int_0^{t}(\beta \sigma)(\cdot,\theta,\tau) d\tau \nonumber\\
& =-(\mathds{1}-T)\int_0^{t}(\beta \sigma)(\cdot,\theta,\tau) d\tau \label{eq: 3.15}.
\end{align}

\section{Scaling}\label{sec: scaling section}
Assume that $\mathcal{X}$ and $\mathcal{T}$ are the macroscopic space and time scales respectively. Let us also consider that the mean run time $\bar{\tau}$ is small compared with the macroscopic time $\mathcal{T}$, i.e., $\varepsilon=\nicefrac{\bar{\tau}}{\mathcal{T}}\ll 1$ where $\varepsilon$ is a small parameter. Suppose further that the concentration $\rho$ is already dimensionless in the sense that it stands for $\rho/\rho_0$ where $\rho_0$ is an averaged value of $\rho$ over $\mathbb{R}^n$. 

The new dimensionless variables are $$t_n=\frac{t}{\mathcal{T}},\ \mathbf{x}_n=\frac{\mathbf{x}}{\mathcal{X}},\ \tau_n=\frac{\tau}{\bar{\tau}}\ \mathrm{and}\ c_n=\frac{c}{s}.$$
We consider the scaling $$t_n=\varepsilon t,\  \mathbf{x}_n=\frac{\varepsilon \mathbf{x}}{s},\  c_n=\varepsilon^{-\gamma} c_0\ \textnormal{and}\  \tau_n=\tau\varepsilon^{\mu},$$ 
for $\mu>0$ and $0<\gamma<1$. Equations \cref{eq: survival} and \cref{eq:beta} become, after substituting the new variables, 
\begin{equation}
\psi_\varepsilon(\cdot,\theta,\tau)=\left( \frac{\tau_0\varepsilon^{\mu}+\varepsilon^{\mu}\tau_1 D^\theta_\varepsilon\rho}{\tau_0\varepsilon^{\mu}+\varepsilon^{\mu}\tau_1 D^\theta_\varepsilon\rho+\tau}\right)^\alpha \label{eq: 3.12}
\end{equation}
and 
\begin{equation}
\beta_\varepsilon(\cdot,\theta,\tau)=\frac{\alpha\varepsilon^{\mu}}{\tau_0\varepsilon^{\mu}+\tau_1\varepsilon^{\mu}D^\theta_\varepsilon\rho+\tau}.
\end{equation}
Here $
D^\theta_\varepsilon\rho=\varepsilon\partial_t \rho+\varepsilon^{1-\gamma} c_0\theta\cdot \nabla\rho.
$ The parameters $\mu$ and $\gamma$ will be chosen appropriately in \Cref{sec: scaling}. Note that the scaling chosen here suggests that the macroscopic equation is valid in the scale of the experiments shown in \Cref{fig: msd of cells}.

The scaling of \cref{eq: 3.15} gives 
\begin{equation}
\varepsilon\partial_t\bar{\sigma}+\varepsilon^{1-\gamma}c_0\mathbf{\theta}\cdot \nabla \bar{\sigma}=-(\mathds{1}-T)\int_0^{t}\beta_\varepsilon \sigma d\tau \label{eq: 3.13}.
\end{equation}
Under the appropriate scaling we will pass to the limit when $\varepsilon\rightarrow 0$ and obtain a fractional Patlak-Keller-Segel equation describing the singular limit. 

To do so, we first obtain a conservation equation by integrating \cref{eq: 3.13} over $\theta$ in the whole sphere $S$ and use the conservation of particles, \cref{eq: conservation T}. This gives
\begin{align*}
 \varepsilon\partial_t\frac{1}{|S|}\int_S\bar{\sigma}d\theta+\varepsilon^{1-\gamma}\frac{c_0}{|S|}\nabla\cdot\int_S\theta\bar{\sigma}d\theta & =0.
\end{align*}
The mean direction $\bar{w}=\frac{1}{|S|}\int_S\theta\bar{\sigma}d\theta$ (\Cref{sec: turn_angle_properties}) is calculated in \Cref{sec: scaling} in terms of a new density $\bar{u}=\frac{1}{|S|}\int_S\bar{\sigma} d\theta$. After substituting the mean direction into the conservation equation we will obtain a nonlocal diffusion equation for $\bar{u}$.

\section {Derivation of the turning operator}\label{sec: derivation of turning operator}

In this section we derive the turning operator, given by a kernel $\mathcal{B}$, that describes the behaviour of the individuals. 

We define the density of cells leaving the point $\mathbf{x}$ for all times $\tau$ from $0$ to $t$, also called the escape rate, as
\begin{equation}
i(\mathbf{x},t,\mathbf{\theta})=\int_{0}^{t}\beta(\mathbf{x},t,\mathbf{\theta},\tau)\sigma(\mathbf{x},t,\mathbf{\theta},\tau)d\tau.\label{eq: i}
\end{equation}

Recalling the expression for the running probability \cref{eq: relation suvival and beta} and its relationship to the stopping density function $\varphi$,
\begin{equation}
\varphi(\mathbf{x},t,\theta,\tau)=-\partial_\tau\psi(\mathbf{x},t,\theta,\tau) \label{eq: 3.21} ,
\end{equation}
we can write $\beta$
as 
\[
\beta(\mathbf{x},t,\mathbf{\theta},\tau)=\frac{\varphi(\mathbf{x},t,\theta,\tau)}{\psi(\mathbf{x},t,\theta,\tau)}.
\]

Substituting this expression into \cref{eq: i} and using the solution \cref{eq: solution} obtained from the method of characteristics which is given by
\begin{equation}
\sigma(\mathbf{x},t,\mathbf{\theta},\tau)=\sigma(\mathbf{x}-c\mathbf{\theta}\tau,t-\tau,\mathbf{\theta},0)\psi(\mathbf{x},t,\theta,\tau),\label{eq: sigma}
\end{equation}
we get
\begin{align}
i(\mathbf{x},t,\mathbf{\theta}) & =\int_0^t\varphi(\mathbf{x},t,\theta,\tau)\sigma(\mathbf{x}-c\mathbf{\theta}\tau,t-\tau,\mathbf{\theta},0)d\tau\nonumber\\ & =\int_0^t\varphi(\mathbf{x},t,\theta,t-s)e^{-(t-s) c\theta\cdot\nabla}\sigma(\mathbf{x},s,\theta,0)ds,\label{eq: other i}
\end{align}
by letting $\tau=t-s$. In order to find the Laplace transform of \cref{eq: other i} we expand the term $\varphi$  in a quasi-static approximation by freezing coefficients at $t=t_0$,
\begin{align}
i(\mathbf{x},t, & \theta) =\int_0^t\sum_{k=0}^\infty\frac{(t-t_0)^k}{k!}\partial_t^{(k)}\varphi(\mathbf{x},t_0,\theta,t-s)e^{-(t-s)c\theta\cdot\nabla}\sigma(\mathbf{x},s,\theta,0)ds\nonumber\\ & =\int_0^t\varphi(\mathbf{x},t_0,\theta,t-s)e^{-(t-s)c\theta\cdot\nabla}\sigma(\mathbf{x},s,\theta,0)ds+\mathcal{O}\left( (t-t_0)\varphi'(\mathbf{x},t_0,\theta,t-s)\right).\label{eq: quasistatic approximation}
\end{align}
We later let $t_0 \to t$ and keep the leading approximation  in the quasi-static regime.

The Laplace transform of \cref{eq: other i} is 
\begin{align}
\hat{i}(\mathbf{x},\lambda,\theta)=\hat{\varphi}(\mathbf{x},t_0,\theta,\lambda & +c\theta\cdot\nabla)\Big|_{t_0=t}\hat{\sigma}(\mathbf{x},\lambda,\theta,0)\nonumber\\ & +\mathcal{O}\left( (t-t_0)(\lambda+c\theta\cdot\nabla)\hat{\varphi}(\mathbf{x},t_0,\theta,\lambda+c\theta\cdot\nabla)\right) .\label{eq: i transformed}
\end{align}

On the other hand, using the definition of $\bar{\sigma}$ given in \cref{eq: sigma_bar} we also have
\begin{align}
\bar{\sigma}(\mathbf{x},t,\mathbf{\theta}) & =\int_0^t\sigma(\mathbf{x}-c\mathbf{\theta}\tau,t-\tau,\mathbf{\theta},0)\psi(\mathbf{x},t,\theta,\tau)d\tau\nonumber\\ & = \int_0^te^{-(t-s)c\theta\cdot\nabla}\sigma(\mathbf{x},s,\theta,0)\psi(\mathbf{x},t,\theta,t-s)ds.\label{eq: sigma_proof}
\end{align}
Following the same approximation as in \cref{eq: quasistatic approximation}, we obtain the Laplace transform of $\bar{\sigma}$ as follows
\begin{align}
\hat{\bar{\sigma}}(\mathbf{x},\lambda,\theta)=\hat{\sigma}(\mathbf{x},\lambda,\theta,0)\hat{\psi}(\mathbf{x},t_0,\theta,\lambda&+c\theta\cdot\nabla)\Big|_{t_0=t}\nonumber \\ & +\mathcal{O}\left( (t-t_0)(\lambda+c\theta\cdot\nabla)\hat{\psi}(\mathbf{x},t_0,\theta,\lambda+c\theta\cdot\nabla)\right) .\label{eq: sigma transformed}
\end{align}
Finally, from \cref{eq: i transformed} and \cref{eq: sigma transformed} we get
\begin{equation}
\hat{i}(\mathbf{x},\lambda,\theta)=\hat{\mathcal{B}}(\mathbf{x},t,\theta,\lambda+c\theta\cdot\nabla)\hat{\bar{\sigma}}(\mathbf{x},\lambda,\theta)+\textnormal{l.o.t.},\label{eq: final i transformed}
\end{equation}
where we neglect the lower order terms and $\mathcal{B}$ denotes the turning operator defined as 
\begin{equation}
\hat{\mathcal{B}}(\mathbf{x},t,\theta,\lambda+ c\theta\cdot\nabla)=\frac{\hat{\varphi}(\mathbf{x},t,\theta,\lambda+c\theta\cdot\nabla)}{\hat{\psi}(\mathbf{x},t,\theta,\lambda+c\theta\cdot\nabla)}+\textnormal{l.o.t.}.\label{eq:kernel}
\end{equation}
Applying the inverse Laplace transform to \cref{eq: final i transformed} we have
\begin{equation}
i(\mathbf{x},t,\mathbf{\theta})=  \int_{0}^{t}\mathcal{B}(\mathbf{x},t,\theta,t-s)\bar{\sigma}(\mathbf{x}-c\theta(t-s),s,\theta)ds. \label{eq: 3.14}
\end{equation}

Next we find $\hat{\psi}_\varepsilon(\mathbf{x},t,\theta,\lambda)$ and $\hat{\varphi}_\varepsilon(\mathbf{x},t,\theta,\lambda)$ in order to obtain an explicit form for $\hat{\mathcal{B}}_\varepsilon$. The subscript $\varepsilon$ denotes that these quantities are scaled as indicated in \Cref{sec: scaling section}.
For $a=\tau_0(\rho)\varepsilon^{\mu}+\tau_1(\rho)\varepsilon^{\mu}D^\theta_\varepsilon\rho$, the Laplace transform of $\psi_\varepsilon$ given in \cref{eq: 3.12} is
\begin{equation}
\hat{\psi}_\varepsilon(\mathbf{x},t,\theta,\lambda)=a^\alpha \lambda^{\alpha-1} e^{a\lambda}\Gamma(-\alpha+1,a\lambda),\label{eq: psi bar}
\end{equation}
in the quasi-static approximation that $D^\theta_\varepsilon\rho$ varies slowly along a run.
Using the following asymptotic expansion for the incomplete Gamma function  
\begin{align}
\Gamma(b,z) & = \Gamma(b)\left( 1-z^{b}e^{-z}\sum_{k=0}^{\infty}\frac{z^k}{\Gamma(b+k+1)}\right),\label{eq: 3.23}
\end{align}
where $b$ is positive non-integer \cite{NIST:DLMF}, and recalling that $b\Gamma(b)=\Gamma(b+1)$, we can rewrite the expression \cref{eq: psi bar} as
\[
\hat{\psi}_\varepsilon(\mathbf{x},t,\theta,\lambda)=-\frac{a}{1-\alpha}-\frac{a^{2}\lambda}{(1-\alpha)(2-\alpha)}+a^{\alpha}\lambda^{\alpha-1}\Gamma(-\alpha+1)+\mathcal{O}(a^3\lambda^2).
\]
 Note that in the above we have considered that $e^{a\lambda}=1+\mathcal{O}(a\lambda)$.
 
 To simplify notation, let us define the quantities
 \begin{align*}
     \zeta=-\frac{a}{1-\alpha}, & & \vartheta=\frac{a^2}{(1-\alpha)(2-\alpha)}, & & \eta=a^\alpha \Gamma(-\alpha+1),
 \end{align*}
 which are respectively of order $a$, $a^2$, and $a^\alpha$. Then $\hat{\psi}_\varepsilon$ is 
 \begin{equation}
 \hat{\psi}_\varepsilon(\mathbf{x},t,\theta,\lambda)=\zeta-\vartheta\lambda+\eta\lambda^{\alpha-1}+\mathcal{O}(a^3\lambda^2).\label{eq: psi hat}
 \end{equation}
From a geometric expansion in $a \lambda \neq 0$ and the binomial theorem we have
\begin{align}
\left(\hat{\psi}_\varepsilon(\mathbf{x},t,\theta,\lambda)\right)^{-1} & = \frac{1}{\zeta}\sum_{k=0}^\infty\left(\frac{\vartheta}{\zeta}\lambda-\frac{\eta}{\zeta}\lambda^{\alpha-1} \right)^k\nonumber\\ & =\frac{1}{\zeta}+\frac{1}{\zeta}\sum_{k=2}^{\frac{1}{\alpha-1}}\left(\frac{\eta}{\zeta}\right)^k\lambda^{k(\alpha-1)}-\frac{\eta}{\zeta^2}\lambda^{\alpha-1}+\frac{\vartheta}{\zeta^2}\lambda+\mathcal{O}\left(a^{-1}(a\lambda)^\alpha \right)\label{eq: 3.26}
\end{align}
 for $ k\in\mathbb{N}$ and 
 $
 \left|\vartheta\lambda-\eta\lambda^{\alpha-1}\right|<\zeta
 $, since the left hand side is of higher order in $a \lambda$. The terms in the sum over $k$ will eventually be of lower order in the scaling parameter $\varepsilon$ in \Cref{sec: scaling}. We neglect them in the following. 
 
Solving \cref{eq: 3.21} we obtain
\begin{equation}
\varphi_\varepsilon(\mathbf{x},t,\theta,\tau)=\frac{\alpha a^\alpha}{(a+\tau)^{\alpha+1}},\label{eq:3.22}
\end{equation}
and the Laplace transform of \cref{eq:3.22} is 
\[
\hat{\varphi}_\varepsilon(\mathbf{x},t,\theta,\lambda) =\alpha (a\lambda)^\alpha\Gamma(-\alpha,a\lambda)e^{a\lambda}.
\]

Again using the expansion for the incomplete Gamma function (\ref{eq: 3.23}) we see that
\begin{equation}
\hat{\varphi}_\varepsilon(\mathbf{x},t,\theta,\lambda)=1+\frac{a\lambda}{1-\alpha}+\mathcal{O}(a^\alpha\lambda^\alpha).\label{eq: laplace_phi}
\end{equation}

As a consequence, from \cref{eq: 3.26} and \cref{eq: laplace_phi} we conclude
\begin{equation}
     \frac{\hat{\varphi}_\varepsilon(\mathbf{x},t,\theta,\lambda)}{\hat{\psi}_\varepsilon(\mathbf{x},t,\theta,\lambda)}=\frac{\alpha-1}{a}-\frac{\lambda}{2-\alpha}-a^{\alpha-2}\lambda^{\alpha-1}(\alpha-1)^2\Gamma(-\alpha+1)+\mathcal{O}(a^{\alpha-1}\lambda^\alpha),\label{eq: B final}
 \end{equation}
up to lower order terms in $a \lambda$.

\subsection {Fractional diffusion equation}

Using the form of $\beta$ obtained in the previous part, the scaled model equation \cref{eq: 3.13} takes the form
\begin{align*}
    \varepsilon\partial_t\bar{\sigma}+\varepsilon^{1-\gamma}c_0\theta \cdot\nabla\bar{\sigma} & =-(\mathds{1}-T)\int_{0}^{t}\mathcal{B}_\varepsilon(\mathbf{x},t,\theta,t-s)\bar{\sigma}(\mathbf{x}- c\theta(t-s),s,\theta)ds. 
\end{align*}
Computing the Laplace transform of the above expression, we obtain
\begin{equation}
    [\varepsilon\lambda+\varepsilon^{1-\gamma} c_0\theta\cdot\nabla]\hat{\bar{\sigma}}(\mathbf{x},\lambda,\mathbf{\theta})-\varepsilon\bar{\sigma}_0(\mathbf{x},\mathbf{\theta})=-(\mathds{1}-T)\hat{\mathcal{B}}_\varepsilon(\mathbf{x},t,\mathbf{\theta},\varepsilon\lambda+\varepsilon^{1-\gamma} c_0\theta\cdot\nabla)\hat{\bar{\sigma}}(\mathbf{x},\lambda,\theta).
\end{equation}
We substitute $\mathcal{B}_\varepsilon(\mathbf{x},t,\theta,\varepsilon\lambda+\varepsilon^{1-\gamma}c_0\theta\cdot\nabla)$ with \cref{eq:kernel} in the above expression and use \cref{eq: 3.26} and \cref{eq: laplace_phi} to obtain  
\begin{align}
[\varepsilon\lambda &+\varepsilon^{1-\gamma} c_0\mathbf{\theta} \cdot\nabla] \hat{\bar{\sigma}}(\mathbf{x},\lambda,\mathbf{\theta}) -\varepsilon\bar{\sigma}_0(\mathbf{x},\mathbf{\theta}) =-(\mathds{1}-T)\Bigl[\frac{1}{\zeta} +\frac{\vartheta}{\zeta^2}(\varepsilon\lambda+\varepsilon^{1-\gamma}c_0\mathbf{\theta}\cdot\nabla)\nonumber\\ & -\frac{\eta}{\zeta^2}(\varepsilon\lambda+\varepsilon^{1-\gamma}c_0\mathbf{\theta}\cdot\nabla)^{\alpha-1} +\mathcal{O}(a^{\alpha-1}\lambda^\alpha)\Bigr]\hat{\varphi}_\varepsilon(\mathbf{x},t,\theta,\varepsilon\lambda+\varepsilon^{1-\gamma}c_0\mathbf{\theta}\cdot\nabla)\hat{\bar{\sigma}}(\mathbf{x},\lambda,\mathbf{\theta}).
\end{align}
Recalling that $1-\gamma<1$ we find that to leading order in $\varepsilon$
\begin{align*}
(\varepsilon\lambda+ \varepsilon^{1-\gamma}c_0\mathbf{\theta}\cdot\nabla)^{\alpha-1} & =( \varepsilon^{1-\gamma}c_0\mathbf{\theta}\cdot\nabla)^{\alpha-1}+\mathcal{O}\left(\varepsilon^{1+(\alpha-1)(1-\gamma)}\right).
\end{align*}
Hence,
\begin{align}
[\varepsilon\lambda &+\varepsilon^{1-\gamma} c_0\mathbf{\theta}\cdot\nabla] \hat{\bar{\sigma}}(\mathbf{x},\lambda,\mathbf{\theta})-\varepsilon\bar{\sigma}_0(\mathbf{x},\mathbf{\theta})=-(\mathds{1}-T)\Bigl[\frac{1}{\zeta}+\frac{\vartheta}{\zeta^2} \varepsilon^{1-\gamma}c_0\mathbf{\theta}\cdot\nabla\nonumber\\ & -\frac{\eta}{\zeta^2}( \varepsilon^{1-\gamma}c_0\mathbf{\theta}\cdot\nabla)^{\alpha-1}+\mathcal{O}\left(\varepsilon^{1+(\alpha-1)(-\mu-\gamma+1)}\right)\Bigr]\hat{\varphi}_\varepsilon(\mathbf{x},t,\theta,\varepsilon^{1-\gamma}c_0\mathbf{\theta}\cdot\nabla)\hat{\bar{\sigma}}(\mathbf{x},\lambda,\mathbf{\theta}).\label{eq: laplace_fourier}
\end{align}

Transforming \cref{eq: laplace_fourier} back to the time domain, we conclude
\begin{align}
    \varepsilon\partial_t\bar{\sigma}+\varepsilon^{1-\gamma}c_0\theta \cdot\nabla\bar{\sigma} =-(\mathds{1}-T)\mathcal{T}_\varepsilon\bar{\sigma},\label{eq: last}
\end{align}
where to leading order
\[
\mathcal{T}_\varepsilon=\frac{\hat{\varphi}_\varepsilon(\mathbf{x},t,\theta,\varepsilon^{1-\gamma}c_0\theta\cdot\nabla)}{\hat{\psi}_\varepsilon(\mathbf{x},t,\theta,\varepsilon^{1-\gamma}c_0\theta\cdot\nabla)}.
\]

\section{Scaling analysis}\label{sec: scaling}
We expand  $\bar{\sigma}_\varepsilon$ using the eigenfunction representation from \Cref{sec: turn_angle_properties}: 
\[
\bar{\sigma}_\varepsilon=\frac{1}{|S|}(\bar{u}+\varepsilon^{\kappa}n\mathbf{\theta}\cdot\bar{w})+\textnormal{l.o.t.},
\]
where $\kappa>0$. Note that the lower order terms are orthogonal to all linear polynomials in $\theta$. Substituting the expansion into \Cref{eq: last},
\begin{equation}
\frac{\varepsilon}{|S|}\partial_t\left( \bar{u}+\varepsilon^{\kappa}n\theta\cdot \bar{w}\right)+ \frac{\varepsilon^{1-\gamma} c_0}{|S|}\theta\cdot\nabla\left(\bar{u}+\varepsilon^{\kappa}n\theta\cdot \bar{w} \right)=-\frac{1}{|S|}(\mathds{1}-T)\mathcal{T}_\varepsilon\left(\bar{u}+\varepsilon^{\kappa}n\theta\cdot \bar{w}\right), \label{eq: long}
\end{equation}
up to lower order terms.
By integrating over $\mathbf{\theta}$ and recalling \cref{eq: conservation T} (\Cref{sec: turn_angle_properties}), as well as the above-mentioned orthogonality, we find the macroscopic conservation equation
\begin{equation}
\varepsilon\partial_t\bar{u}+\varepsilon^{\kappa+1-\gamma }c_0n\nabla\cdot\bar{w}=0.\label{eq: conservation}
\end{equation}
This equation is nontrivial only for $\kappa=\gamma$, so that
\[
\bar{\sigma}_\varepsilon=\frac{1}{|S|}(\bar{u}+\varepsilon^{\gamma}n\mathbf{\theta}\cdot\bar{w})+\textnormal{l.o.t.}.
\]
To obtain an equation for the mean direction $\bar{w}$, we multiply \cref{eq: long} by $\theta$ and integrate over the whole sphere $S$:
\begin{align}
    n\varepsilon^{\gamma+1}\partial_t\bar{w}+\varepsilon^{1-\gamma} c_0\nabla \bar{u}=-\frac{1}{|S|}\int_S\theta(\mathds{1}-T)\mathcal{T}_\varepsilon\left( \bar{u}+\varepsilon^{\gamma}n\theta\cdot \bar{w}\right)d\theta.\label{eq: important equation}
\end{align}

Using \Cref{eq: important equation} and the appropriate values for $\mu$ and $\gamma$, we get an expression for $\bar{w}$ which, on substitution into the conservation equation \cref{eq: conservation}, leads to the fractional Patlak-Keller-Segel equation.

To see this, we first determine $\mathcal{T}_\varepsilon$. Considering \cref{eq: B final} we have,
\begin{align}
    \mathcal{T}_\varepsilon =\frac{\alpha-1}{a}  -\frac{\varepsilon^{1-\gamma}c_0}{2-\alpha}(\theta\cdot\nabla) -a^{\alpha-2}(\alpha-1)^2\Gamma(&-\alpha+1)(\varepsilon^{1-\gamma}c_0\theta\cdot\nabla)^{\alpha-1} \nonumber\\ &+\mathcal{O}(\varepsilon^{\mu-\gamma+1}).\label{eq: almost final}
\end{align}
We notice that $D^\theta_\varepsilon\rho=\varepsilon\partial_t\rho+\varepsilon^{1-\gamma}c_0\theta\cdot\nabla\rho \simeq \varepsilon^{1-\gamma}c_0\theta\cdot\nabla\rho$ since $1-\gamma<1$. Then, expanding the term
$$
a^{\alpha-2}=(\tau_0\varepsilon^{\mu})^{\alpha-2}\left(1+\frac{\tau_1}{\tau_0}\varepsilon^{1-\gamma}c_0\theta\cdot\nabla\rho  \right)^{\alpha-2}
$$
using a binomial expansion, we find 
\[
a^{\alpha-2}=(\tau_0\varepsilon^{\mu})^{\alpha-2}\left( 1+\varepsilon^{1-\gamma}(\alpha-2)\frac{\tau_1}{\tau_0}c_0\theta\cdot\nabla\rho+\mathcal{O}\left(\varepsilon^{2(1-\gamma)}\right) \right).
\]
Similarly we can write 
\[
a^{-1}=\frac{\varepsilon^{-\mu}}{\tau_0}\left(1-\varepsilon^{1-\gamma}\frac{\tau_1}{\tau_0}c_0\theta\cdot\nabla\rho+\mathcal{O}\left(\varepsilon^{2(1-\gamma)}\right)  \right).
\]
Therefore, the operator in \cref{eq: almost final} becomes
\begin{equation}
\begin{aligned}
    \mathcal{T}_\varepsilon & =\Bigl[\frac{\varepsilon^{-\mu}}{\tau_0}(\alpha-1)-\frac{\tau_1}{\tau_0^2}(\alpha-1)\varepsilon^{-\mu-\gamma+1} c_0\theta\cdot\nabla\rho-\frac{\varepsilon^{1-\gamma}c_0}{2-\alpha}\theta\cdot\nabla\\ & + \left(-\tau_0^{\alpha-2}\varepsilon^{\mu(\alpha-2)+(1-\gamma)(\alpha-1)}+\tau_0^{\alpha-3}\tau_1(2-\alpha)\varepsilon^{\mu(\alpha-2)+\alpha(1-\gamma)}c_0\theta\cdot\nabla\rho\right)\\ &(1-\alpha)^2\Gamma(-\alpha+1)c_0^{\alpha-1}(\theta\cdot\nabla)^{\alpha-1}\Bigr]+\mathcal{O}\left(\varepsilon^{\mu-\gamma+1}\right).\label{eq: important-important}
\end{aligned}
\end{equation}
The physically relevant scaling regime involves transport in the equation for $\bar{w}$. For $-\mu-\gamma+1=-\mu+\gamma$ we obtain $\gamma=1/2$, and therefore $\mu=\frac{2-\alpha}{2(\alpha-1)}.$ This scaling leads to 
\begin{align*}
    \mathcal{T}_\varepsilon(\bar{u} & +\varepsilon^{-\gamma}n\theta\cdot\bar{w}) = \frac{\varepsilon^{\frac{\alpha-2}{2(\alpha-1)}}}{\tau_0}(\alpha-1)\bar{u}-\frac{\tau_1}{\tau_0^2}(\alpha-1)\varepsilon^{\frac{\alpha-2}{2(\alpha-1)}+\frac{1}{2}}c_0(\theta\cdot\nabla\rho)\bar{u}\\ & -\tau_0^{\alpha-2}(1-\alpha)^2\Gamma(-\alpha+1)\varepsilon^{\frac{\alpha-2}{2(\alpha-1)}+\frac{1}{2}}c_0^{\alpha-1}(\theta\cdot\nabla)^{\alpha-1}\bar{u} +\frac{\varepsilon^{\frac{\alpha-2}{2(\alpha-1)}+\frac{1}{2}}}{\tau_0}(\alpha-1)n\theta\cdot\bar{w}\\ & +\mathcal{O}(\varepsilon^{\frac{\alpha-2}{2(\alpha-1)}+1}).
\end{align*}
We now compare the leading powers of $\varepsilon$ in \Cref{eq: important equation}. For the coefficient of the leading term $\varepsilon^{\frac{\alpha-2}{2(\alpha-1)}}$ we find
\begin{align}
 0=-\frac{1}{|S|}\int_S\theta(\mathds{1}-T)\frac{\alpha-1}{\tau_0}\bar{u}d\theta, \label{eq: jodedera}
\end{align}
while the subleading term is of order $\varepsilon^{\frac{2\alpha-3}{2(\alpha-1)}}$ with coefficients
\begin{align} 
&0=-\frac{1}{|S|}\int_S\theta(\mathds{1}-T)\Bigl[-\frac{\tau_1}{\tau_0^2}(\alpha-1)c_0(\theta\cdot\nabla\rho)\bar{u}\nonumber\\ &\qquad -\tau_0^{\alpha-2}(1-\alpha)^2\Gamma(-\alpha+1)c_0^{\alpha-1}(\theta\cdot\nabla)^{\alpha-1}\bar{u} +\frac{\alpha-1}{\tau_0}n\theta\cdot\bar{w}\Bigr]d\theta.\label{eq: casi final}
\end{align}
We see from  \cref{eq: jodedera}:
\begin{equation}
-\frac{1}{|S|}\int_S\theta(\mathds{1}-T)\frac{\alpha-1}{\tau_0}\bar{u}d\theta=-\frac{1}{|S|}\frac{\alpha-1}{\tau_0}\bar{u}\int_S\theta(\mathds{1}-T)d\theta=0,
\end{equation}
due to the conservation condition \cref{eq: conservation T}. Similarly, \cref{eq: casi final} becomes, using the representation of $T$ in terms of its eigenfunctions as in \cref{eq: operator T} in \Cref{app: fractional},
\begin{align}
0 &= -\frac{1}{|S|}\int_S\theta\Bigl( \frac{\tau_1}{\tau_0^2}(\alpha-1)c_0(\theta\cdot\nabla\rho)\bar{u}(\nu_1-1)-\tau_0^{\alpha-2}(1-\alpha)^2\Gamma(-\alpha+1)(c_0\theta\cdot\nabla)^{\alpha-1}\bar{u}\nonumber\\ &\hspace{-0.3cm} +\tau_0^{\alpha-2}(1-\alpha)^2\Gamma(-\alpha+1)c_0^{\alpha-1}\left(\frac{\mathds{D}^{\alpha-1}}{|S|}+\frac{n^2\nu_1}{|S|}\theta\cdot\nabla^{\alpha-1}\right)\bar{u}-\frac{\alpha-1}{\tau_0}n\theta\cdot\bar{w}(\nu_1-1)\Bigr)d\theta \nonumber\\ &= -\frac{\tau_1}{\tau_0^2}(\alpha-1)c_0\bar{u}(\nu_1-1)\nabla\rho-\tau_0^{\alpha-2}(1-\alpha)^2\Gamma(-\alpha+1)c_0^{\alpha-1}\nabla^{\alpha-1}\bar{u}\left(\frac{n^2\nu_1}{|S|}-1\right) \nonumber\\ &\hspace{-0.3cm} +\frac{\alpha-1}{\tau_0}n\bar{w}(\nu_1-1).\nonumber
\end{align}

\begin{figure}[htbp]
\centering
    \includegraphics[scale=0.5]{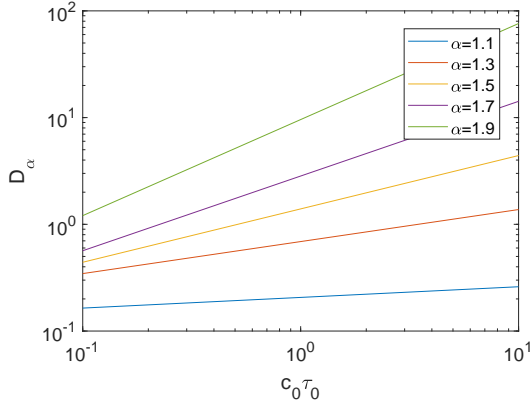}
    \caption{Diffusion coefficient $D_\alpha$ for different values of $c_0\tau_0$, where the remaining parameters have being left fixed.}\label{fig: alphas}
\end{figure}

We can solve this for the mean flux $c_0\bar{w}$, which is given by 
\begin{equation}
    c_0\bar{w}=\frac{\tau_1}{n\tau_0}c_0^2\bar{u}\nabla\rho+\frac{\pi\tau_0^{\alpha-1}(\alpha-1)}{\sin(\pi\alpha)\Gamma(\alpha)}\frac{(n^2\nu_1-|S|)}{n|S|(\nu_1-1)}c_0^\alpha\nabla^{\alpha-1}\bar{u}\ ,
\end{equation}
where we have used $\Gamma(-\alpha+1)=\frac{\pi}{\sin(\pi\alpha)\Gamma(\alpha)}$.
The conservation equation \cref{eq: conservation} can therefore be written as 
\begin{equation}
\partial_t\bar{u} =c_0\nabla\cdot(D_\alpha\nabla^{\alpha-1}\bar{u}-\chi\bar{u}\nabla\rho)\label{eq:final equation}
\end{equation}
for 
\[
D_\alpha=-\frac{\pi(\tau_0c_0)^{\alpha-1}(\alpha-1)}{\sin(\pi\alpha)\Gamma(\alpha)}\frac{(n^2\nu_1-|S|)}{|S|(\nu_1-1)}\ \textnormal{and}\  \chi=\frac{\tau_1c_0}{\tau_0}.
\]
Note that $D_\alpha>0$ since $\sin(\pi\alpha)<0$ for $1<\alpha<2$. \Cref{fig: alphas} shows the behaviour of the diffusion coefficient for different values of $c_0\tau_0$, depending on $\alpha$.

Note further that using a Cattaneo approximation to approximate the effective contribution of higher order terms leads to an additional diffusive term in \cref{eq:final equation}. However, the coefficient of this term turns out to be of lower order in the scaling variable $\varepsilon$ and hence can be neglected.

\section{Conclusions \& Outlook}
In this paper we have derived effective macroscopic diffusion equations for organisms with long range behaviour, in the presence of some chemoattractant or nutrient. Beginning with a microscopic model in which run time distributions follow a power-law as observed, for example, for E. coli and Dd at low nutrient concentrations, we obtain the form of the scattering operator and the resulting kinetic equation. The fractional Patlak-Keller-Segel system \cref{eq:final equation} emerges in a realistic hyperbolic limit.

Unlike in \cite{bellouquid2016kinetic}, where the authors derived a similar fractional diffusion equation starting from a kinetic equation, our approach starts from a model for the individual organisms, reflecting the experimentally observed movement patterns. This model can subsequently be made concrete in a wide range of different biological contexts.

Our discussion in this article focused on organisms in an unbounded domain or sufficiently far away from physical boundaries. This would seem reasonable for cells tracked in vitro under the microscope, where the containing disk is multiple orders of magnitude larger than a cell. However, the nonlocality of  \cref{eq:final equation} will lead to a significantly increased influence of the boundary as well as the surrounding environment, compared to standard diffusion \cite{alt1980singular}. 
The actual interaction between an organism and the boundary is expected to vary considerably according to the organism and the nature of the system (for example between an experiment and a natural environment). Any meaningful discussion of boundary conditions would therefore have to focus on the context of modelling a particular biological system.
For an example of an anomalous effective diffusion equation with Dirichlet boundary conditions, see \cite{cesbron2016anomalous}. A more detailed discussion of relevant boundary conditions for fractional Patlak-Keller-Segel equations will be considered in future work.
  
More generally, it will certainly be fruitful to tailor the model to particular biological systems. For example in the case of \textit{E. coli}, the stopping probability could be specifically linked to molecular components (e.g. CheR) which enter as internal variables. We refer to work in this direction by Perthame et.~al.~\cite{perthame2016derivation} for the run-and-tumble of bacteria including a biochemical pathway. For a more detailed discussion of modelling bacterial chemotaxis including internal variables we refer to \cite{xue2009individual} and references therein.  

In the current paper our assumptions have been largely motivated by the motion of \textit{E. coli}, which offers an opportunistic case study due to its well characterised behaviour. While the results give some insight into the expected equations for other cells or organisms, extending to such systems in a more meaningful way would require re-evaluation of the core assumptions. For example, eukaryotic cells such as {\em Dd} or immune cells can be large enough to directly sense a spatial gradient, so that the turning distribution  is potentially biased with respect to the chemoattractant gradient. Nevertheless, moving to such cell types provides an exciting focus for applications, with T cell movement in the central nervous system (CNS) being one such example. CNS resistance to the encephalitis causing pathogen {\em Toxoplasma gondii} demands that patrolling T cells locate potentially sparsely distributed infection sites. Data in [19] suggest that the immune cells optimise searching via a generalised L\'{e}vy walk involving fixed velocity straight runs with distances randomly chosen from a L\'{e}vy distribution, as in our above assumptions, but also interspersed by pauses that are also drawn from a L\'{e}vy distribution. Adapting the model to this system, however, would allow us to quantitatively investigate how this behaviour increases searching efficiency.
  
Evidence for L\'{e}vy walk type behaviour often seem to arise under very specific conditions: for example, the presence or absence of food or chemoattractant in organism movement. Modelling-wise, this suggests that generalized running probabilities could include switches from a power-law type distribution to exponential law, where the control is specifically mediated by the chemical concentration and/or gradient.
Such \enquote{switching} behaviour between local and nonlocal search has been suggested in \cite{li2008persistent}. Its accurate mathematical modelling remains an open challenge.
  
Furthermore, the impact of interactions among individuals in swarming bacteria appears to be related to the emergence of superdiffusion: See \cite{fedotov2017emergence} for a first work in this direction. Based on experimental results in \cite{ariel2015swarming}, the authors show that L\'{e}vy walks can emerge as a cooperative effect without assuming a power-law distribution of run distances. Nevertheless, the appearance of L\'{e}vy walks in the case of systems of interacting self-propelled particles remains unknown.

Also, mathematically, the analysis of equations of the form \cref{eq:final equation} is of high current interest and has been extensively studied. In \cite{burczak2016suppression} the authors proved existence of global in time 
solutions for certain initial data, for the case of a fractional parabolic-elliptic 
Keller-Segel equation. 
Travelling wave solutions in the case of equations like \cref{eq:final equation} are expected to lead to new phenomena and in particular could be expected to speed up with time, see for instance \cite{cabre2013influence}. In the absence of processes such as proliferation, travelling bands of bacteria dissipate over time in classical Keller-Segel equations, unless bacteria are given \enquote{extreme} sensitivity responses \cite{xue2011travelling}: this dissipation occurs as individuals drop away from the main band and lose contact with the chemoattractant. It is tempting to speculate that giving such \enquote{lost} bacteria an improved searching through fractional diffusion may allow them to reconnect with the main band. 

Other aspects of chemotaxis equations in general, such as pattern formation, are intensely studied, see for example \cite{bellomo2015toward}, and relevant for biological and ecological applications \cite{bullock2017synthesis}.
Numerical investigations should allow us to address some of the previous questions about the dynamics, pattern formation and travelling wave solutions in realistic systems. This is the topic of ongoing work.

\appendix
\section{Turn angle operator}\label{sec: turn_angle_properties}
 
This section recalls some basic spectral properties of the turn angle operator $T$ defined in \cref{eq: turn angle operator}. Crucially, its kernel $k(\mathbf{x},t,\mathbf{\theta};\mathbf{\eta}) =  \ell(\mathbf{x},t,|\eta-\theta|)$ only depends on the distance $|\eta-\theta|$: 
\begin{align*}
T\phi(\eta) & =\int_S k(\mathbf{x},t,\mathbf{\theta};\mathbf{\eta})\phi(\theta)d\theta = \int_S \ell(\mathbf{x},t,|\eta-\theta|)\phi(\theta)d\theta.
\end{align*}
Because $\ell$ is a probability distribution, it is normalized to $\int_S \ell(\mathbf{x},t,|\theta-e_1|)d\theta=1$,  where $e_1 = (1,0,\dots, 0)$. We immediately observe
\begin{align}
    \int_S (\mathds{1}-T)\phi d\theta=0\label{eq: conservation T}
\end{align}
for all $\phi\in L^2(S)$. Biologically, \cref{eq: conservation T} corresponds to the conservation of the number of organisms in the tumbling phase.

We also require some more detailed information about the spectrum of $T$. Recall that in $n$-dimensions, the surface area of the unit sphere $S$ is given by \[
|S|=\begin{cases}
\frac{2\pi^{\nicefrac{n}{2}}}{\Gamma\left(\frac{n}{2}\right)}, & \textnormal{for $n$ even},\\
\frac{\pi^{\nicefrac{n}{2}}}{\Gamma\left(\frac{n}{2}+1\right)}, & \textnormal{for $n$ odd}.
\end{cases}
\]
\begin{lemma}\label{lem: eigenfunctions}
Assume that $\ell$ is continuous. Then $T$ is a symmetric compact operator. In particular, there exists an orthonormal basis of $L^2(S)$ consisting of eigenfunctions of $T$.\\ With $\mathbf{\theta}=(\theta_0,\theta_1,...,\theta_{n-1}) \in S$, we have
\begin{equation}
\begin{aligned}\phi_{0}(\theta) & =\frac{1}{|S|} &  & \text{is an eigenfunction to the eigenvalue} &  & \nu_{0}=1,\\
\phi_{1}^j(\theta) & =\frac{n\theta_j}{|S|} &  & \text{are eigenfunctions to the eigenvalue} &  & \nu_{1}=\int_{S}\ell(\cdot,|\eta-1|)\eta_{1}d\eta<1. \label{eq: eigen}
\end{aligned}
\end{equation}
Any function $\bar{\sigma}\in L^2(\mathds{R}^n\times \mathds{R}^+\times S)$ admits a unique decomposition 
\begin{equation}
\bar{\sigma}=\frac{1}{|S|}\left(\bar{u}+n\mathbf{\theta}\cdot \bar{w} \right)+\hat{z},\label{eq: real_eigen}
\end{equation}
where $\hat{z}$ is orthogonal to all linear polynomials in $\theta$. Explicitly,
\begin{equation*}
\bar{u}(\mathbf{x},t)=\int_S\bar{\sigma}(\mathbf{x},t,\mathbf{\theta})\phi_0(\theta) d\theta,\ \bar{w}^j(\mathbf{x},t)=\int_S \bar{\sigma}(\mathbf{x},t,\mathbf{\theta})\phi_1^j(\theta) d\theta,
\end{equation*}
and  $\bar{w} =  (\bar{w}^1, \dots, \bar{w}^n)$.
\end{lemma}

We interpret $\bar{u}$ as the density of organisms independent of the direction and $\bar{w}$ as their mean direction.

\section{Fractional operators}\label{app: fractional}
We recall some basic definitions concerning fractional differential operators, as well as their relation to the turning operator $T$.

\begin{definition}\label{def: fractional}
For $s \in (0,2)$ and $f \in C^2(\mathds{R}^n)$ define the fractional gradient of $f$ as
\begin{equation}
\nabla^s f(\mathbf{x})=\frac{1}{|S|}\int_{S}\mathbf{\theta}\mathbf{D}_{\mathbf{\theta}}^s f(\mathbf{x})d\mathbf{\theta}=\frac{1}{|S|}\int_{S}\mathbf{\theta}(\mathbf{\theta}\cdot\nabla)^s f(\mathbf{x})d\mathbf{\theta},\label{eq: fractional derivative}
\end{equation}
where $\mathbf{D}_{\mathbf{\theta}}^s=(\mathbf{\theta}\cdot\nabla)^s$ is the fractional directional derivative of order $s$.
The fractional Laplacian of $f$ is given by
\begin{equation}
\mathds{D}^s f(\mathbf{x})=\frac{1}{|S|}\int_{S}\mathbf{D}^s_\mathbf{\theta}f(\mathbf{x})d\mathbf{\theta}.\label{eq: fractional Laplacian}
\end{equation}
\end{definition}
$\mathds{D}^s$ is associated to $(-\Delta)^{\nicefrac{\alpha}{2}}$ as follows,
\[
\mathds{D}^sf(\mathbf{x})=\varXi_\alpha(-\Delta)^{\nicefrac{\alpha}{2}}
\]
where, in two dimensions, for $1<\alpha<2$,
\[
\varXi_\alpha=-2\sqrt{\pi}\cos\left(\frac{\pi\alpha}{2} \right)\frac{\Gamma\left(\frac{\alpha+1}{2} \right)}{\Gamma\left(\frac{\alpha+2}{2} \right)}.
\]
See \cite{meerschaert2006fractional} and \cite{taylor2016fractional} for further information.

Using  \Cref{lem: eigenfunctions} and the definitions  \cref{eq: fractional derivative} and \cref{eq: fractional Laplacian}, we obtain for sufficiently smooth functions $f$ and $\rho$:
\begin{equation}
\begin{aligned}
T(\mathbf{\theta}\cdot\nabla)f & =\nu_1(\mathbf{\theta}\cdot\nabla)f,\ T(\mathbf{\theta}\cdot\nabla\rho)f=\nu_1(\mathbf{\theta}\cdot\nabla\rho)f,\\  T(\mathbf{\theta\cdot\nabla})^{s} f& \simeq\frac{1}{|S|}\int_S\frac{1}{|S|}(\eta\cdot\nabla)^{s}f\ d\eta+\nu_1\frac{n\theta}{|S|}\int_S\frac{n\eta}{|S|}(\eta\cdot \nabla)^{s}f\ d\eta \\ &=\frac{\mathds{D}^{s}f}{|S|}+\nu_1\frac{n^2\theta}{|S|}\cdot\nabla^{s}f.\label{eq: operator T}
\end{aligned}
\end{equation}

\bibliographystyle{siamplain}
\bibliography{references}
\end{document}